\documentclass[prd,superscriptaddress,amsfonts,amssymb,amsmath,showpacs,twocolumn]{revtex4-2}
\usepackage{bm}
\usepackage{amsfonts}
\usepackage{latexsym}
\usepackage{graphicx}
\usepackage{amsmath}
\usepackage{palatino}
\usepackage{mathpazo}
\usepackage{textcomp}
\usepackage{float}
\usepackage{booktabs}
\usepackage{dcolumn}
\usepackage{booktabs}
\usepackage{multirow}
\usepackage{hyperref}
\usepackage{orcidlink}
\hypersetup{colorlinks,citecolor=blue}
\usepackage{xcolor}

\begin{document}

\color{black}       

\title{ Dark Energy and Cosmic Evolution:  A Study in $f(R,T)$ Gravity}

\author{ N. Myrzakulov\orcidlink{0000-0001-8691-9939}}\email{nmyrzakulov@gmail.com}
\affiliation{L N Gumilyov Eurasian National University, Astana 010008, Kazakhstan}

\author{ S. H. Shekh\orcidlink{0000-0003-4545-1975}}\email{  da\_salim@rediff.com}
\affiliation{L N Gumilyov Eurasian National University, Astana 010008, Kazakhstan}
\affiliation{Department of Mathematics, S.P.M. Science and Gilani Arts, Commerce College, Ghatanji, Yavatmal, \\Maharashtra-445301, India.}

\author{Anirudh Pradhan\orcidlink{0000-0002-1932-8431}}\email{pradhan.anirudh@gmail.com}
\affiliation{Centre for Cosmology, Astrophysics and Space Science (CCASS)
	GLA University, Mathura 281406, Uttar Pradesh, India}
	
\author{Archana  Dixit\orcidlink{0000-0003-4285-4162}}\email{archana.ibs.maths@gmail.com}
\affiliation{Department of Mathematics, Gurugram University, Gurugram, Harayana, India}
\begin{abstract}
\textbf{Abstract:}  In the context of $f(R, T)$ gravity theory for the flat Friedmann-Lemaitre–Robertson–Walker (FLRW) model, the accelerating expansion of the universe is investigated using a specific form of the emergent Hubble parameter. Datasets from $H(z)$, Type Ia supernovae (SNIa), and Baryon Acoustic Oscillations (BAO) are used to constrain the model and identify the ideal parameter values in order to evaluate the statistical significance of $f(R, T)$ gravity.
The best-fit parameters are derived by solving the modified Friedmann equations through a MCMC analysis. These parameters are  used to compute the equation of state, statefinders, energy conditions, and the $(\omega-\omega^{'})$ plane. Furthermore, the evolution of kinematic cosmographic parameters is examined. The findings provide significant behavior and features of dark energy models. Our comprehension of the dynamics and evolution of the universe is improved by this study, which also advances our understanding of dark energy and how it shapes the universe.\\
\newline
\textbf{Keywords:} $f(R,T)$ Gravity, $H(z)$ parameterization, Cosmology.
\end{abstract}

\maketitle

\section{Introduction}
The accelerated expansion of the universe has been confirmed through various observations \cite{1,2,3,4,5}. Despite this, \textit{General Relativity} falls short in fully explaining the mechanism behind this cosmic acceleration. Consequently, the expanding universe hypothesis, particularly based on SNeIa data, has been met with some skepticism. To address this, various models have been developed, one approach involves introducing an exotic energy component within the general relativity framework called \textit{dark energy}, with substantial support from theoretical and observational studies \cite{6,7,8}. Another approach modifies the gravitational Lagrangian called \textit{Modified Theories of Gravity}. The Modified theories of gravity  are alternative frameworks designed to extend or revise general relativity to account for phenomena and observations that it cannot fully explain. These theories seek to provide a deeper understanding of the universe's nature and its dynamic behavior. They introduce innovative concepts and mathematical models in an ongoing effort to enhance our knowledge of gravity and its interaction with matter and spacetime. Examples of modified gravity theories include $f(R)$ gravity \cite{9,10,10a}, where $R$ denotes the Ricci scalar; $f(T)$ gravity \cite{11,11a,11b}, with $T$ representing the torsion scalar; and $f(G)$ gravity \cite{12,12a}, based on the Gauss-Bonnet term $G$. Additionally, $f(R,T)$ gravity \cite{13,13a} involves both the Ricci scalar $R$ and the trace of the  energy-momentum tensor $T$, while $f(R,G)$ gravity \cite{14,15,15a,15b} incorporates the Ricci scalar $R$ and the Gauss-Bonnet term $G$. Other notable examples include $f(Q)$ gravity \cite{16,16a,16b,16c,16d,16e,16f}, where $Q$ is the non-metricity scalar, and $f(Q,T)$ gravity \cite{17,17a}, which combines the non-metricity scalar $Q$ with the trace of the energy-momentum tensor $T$. Numerous studies \cite{18,19,20} have investigated $f(R)$ gravity across different theoretical and observational contexts while several researchers \cite{21,22,23,24} have been developed cosmological models in $f(R,T)$ gravity. 
Out of the above-mentioned theories of gravity, we focus on the $f(R,T)$ theory of gravity due to its ability to incorporate the interaction between geometry and matter through the trace of the energy-momentum tensor. This unique feature allows for a more comprehensive exploration of the effects of matter on cosmic evolution, making it a promising framework for addressing the late-time accelerated expansion of the universe. Extensive research has been conducted in this field over the years. This includes investigations into energy condition tests \cite{25}, alternatives to dark matter \cite{26}, cosmological viability tests \cite{27}, the structural configurations of white dwarfs \cite{28}, the Palatini formulation \cite{29}, and the formation of exotic stars \cite{30}. Detailed studies on $f(R,T)$ gravity theory can be found in several significant works \cite{31,32,33,34,35,36,37}.
The action for this theory is defined as \cite{13}:
\begin{equation}\label{1}
	S = \frac{1}{2\kappa} \int f(R,T) \sqrt{-g} \, d^4x + \int \mathcal{L}_m \sqrt{-g} \, d^4x,
\end{equation}

where \(\kappa = 8\pi G / c^4\), \(g\) is the determinant of the metric tensor \(g_{ij}\), and \(\mathcal{L}_m\) is the matter Lagrangian density. To explore the implications of \(f(R,T)\) gravity, several functional forms of \(f(R,T)\) have been proposed. One widely studied form is\cite{13}:
\begin{equation}\label{2}
	f(R,T) = R + 2f(T),
\end{equation}
where \(f(T)\) represents a function of \(T\). In this work, we consider \(f(T) = \lambda T\), where \(\lambda\) is a constant. This choice simplifies the action and allows us to study the interplay between matter and geometry. The field equations for this formulation are derived as:
\begin{widetext}
\begin{equation}\label{3}
	f_R(R,T) R_{ij} - \frac{1}{2} f(R,T) g_{ij} + (g_{ij}\Box - \nabla_i \nabla_j) f_R(R,T) = \kappa T_{ij} - f_T(R,T) T_{ij} - f_T(R,T) \Theta_{ij},
\end{equation}
\end{widetext}

where \(f_R(R,T) = \frac{\partial f(R,T)}{\partial R}\), \(f_T(R,T) = \frac{\partial f(R,T)}{\partial T}\), \(\Box\) is the d'Alembert operator, and \(\Theta_{ij}\) is given by:

\begin{equation}\label{4}
	\Theta_{ij} = g^{lm} \frac{\delta T_{lm}}{\delta g^{ij}}.
\end{equation}

Assuming a perfect fluid, the energy-momentum tensor is expressed as:
\begin{equation}\label{5}
	T_{ij} = (\rho + p) u_i u_j - p g_{ij},
\end{equation}
where \(\rho\) is the energy density, \(p\) is the pressure, and \(u_i\) is the four-velocity vector satisfying \(u^i u_i = 1\). Under this assumption, \(\Theta_{ij}\) reduces to:
\begin{equation}\label{6}
	\Theta_{ij} = -2 T_{ij} - p g_{ij}.
\end{equation}
The field equations simplify to:
\begin{equation}\label{7}
	R_{ij} - \frac{1}{2} R g_{ij} = \kappa T_{ij} + 2f_T T_{ij} + [f(T) + 2p f_T] g_{ij}.
\end{equation}
\section{ Metric and Field Equations}\label{1}
We adopt the spatially flat, homogeneous, and isotropic FLRW metric:
\begin{equation}\label{8}
		ds^2 = dt^2 - a^2(t) \left(dx^2 + dy^2 + dz^2\right),
\end{equation}
where \(a(t)\) is the scale factor, a function of cosmic time \(t\). Substituting this metric into the field equations, we obtain the modified Friedmann equations:
\begin{equation}\label{9}
		3H^2 = (1 + 3\lambda)\rho - \lambda p,
\end{equation}
\begin{equation}\label{10}
		2\dot{H} + 3H^2 = -\lambda\rho + (1 + 3\lambda)p,
\end{equation}
where \(H = \frac{\dot{a}}{a}\) is the Hubble parameter, and \(\dot{H}\) represents the derivative of \(H\) with respect to time.The deceleration parameter \(q\), which characterizes the acceleration of the universe, is defined as:
\begin{equation}\label{11}
		q = -\frac{\ddot{a}a}{\dot{a}^2} = -1 - \frac{\dot{H}}{H^2}.
\end{equation}
The energy density \(\rho\) and pressure \(p\) of the cosmic fluid are expressed in terms of \(H\) and \(\dot{H}\) as:
\begin{equation}\label{12}
		\rho = \frac{1}{(1 + 3\lambda)^2 - \lambda^2} \left[(3 + 6\lambda)H^2 - 2\lambda\dot{H}\right],
\end{equation}
\begin{equation}\label{13}
		p = -\frac{1}{(1 + 3\lambda)^2 - \lambda^2} \left[(3 + 6\lambda)H^2 + 2(1 + 3\lambda)\dot{H}\right].
\end{equation}
These equations provide the foundation for analyzing the cosmological dynamics of the \(f(R,T)\) gravity model. Also, the numerous studies have investigated modifications to the $f(R,T)$ gravity framework using various datasets and methodologies. Nagpal et al. \cite{38} conducted statistical analyses using Hubble, SNeIa, and BAO datasets, adopting the form $f(R,T)=f_1(R)+f_
f(T)$ to represent dynamic vacuum energy in the FLRW cosmological model. Partha et al. \cite{39} explored emergent universe models with bulk viscosity under a flat FLRW, using $f(R,T)=f(R)+f(T)$ alongside Hubble and BAO data. Rudra et al. \cite{40} examined five nonlinear forms of the $f(R,T)$ function, utilizing cosmic chronometer data, BAO, and CMB peaks with the CosmoMC code to propose viable models. Similarly, Pradhan et al. \cite{41} analyzed nonlinear forms of $f(R,T)=R+f(T)$ within the FLRW spacetime using datasets such as Pantheon, BAO, SMALLZ-2014, HST, and PLANCK18, applying the MCMC method. In the following section, we outline several cosmological datasets used to derive observational constraints.
\section{ Cosmological Datasets for Observational Constraints}
 Observational constraints play a vital role in cosmology, as they provide a direct link between theoretical models and real-world observations. By imposing observational constraints, researchers can test the validity of their models, refine their predictions, and underlying physics of the universe. The accuracy and reliability of these constraints are crucial, as they can significantly impact our understanding of the universe's evolution and properties. In this study, we derive observational constraints on the model parameters using a combination of cosmological datasets, including the Supernovae Type Ia datasets, Baryon Acoustic Oscillation and Cosmic Chronometers which provide a comprehensive framework for testing our model against observations. \\
\textit{\textbf{ Supernovae Type Ia}}\\
The SNIa data used in this study are sourced from the Pantheon compilation, consisting of 1048 data points. The chi-square for the supernovae is defined as:
\begin{equation}\label{14}
	\chi^2_{\text{SN}} = (\mu_{\text{obs}} - \mu_{\text{th}})^T C_{\text{Pantheon}}^{-1} (\mu_{\text{obs}} - \mu_{\text{th}}),
\end{equation}
where \(\mu_{\text{obs}}\) is the observed distance modulus, \(\mu_{\text{th}} = 5 \log_{10} \left[c D_L / H_0\right] + 25\), and \(D_L(z)\) is the luminosity distance.\\
\textit{\textbf{Baryon Acoustic Oscillation}}\\
The BAO data are crucial for determining the angular diameter distance \(D_A(z)\) and the spherically averaged distance \(D_V(z)\). These are related to the Hubble parameter \(H(z)\) as:
\begin{equation}\label{15}
	D_V(z) = \left[(1 + z)^2 D_A^2(z) \frac{c z}{H(z)}\right]^{1/3}.
\end{equation}
The total chi-square for BAO is calculated as:
\begin{equation}\label{16}
	\chi^2_{\text{BAO}}(\eta,\beta,\gamma,h) = \sum \chi^2_i,
\end{equation}
where the summation runs over all BAO datasets.\\
\textit{\textbf{Hubble Parameter Data}}\\
The Hubble parameter \(H(z)\) is directly used to constrain the model parameters. The chi-square for the Hubble data is:
\begin{equation}\label{17}
	\chi^2_{H(z)}(\eta,\beta,\gamma,h) = \sum_{i=1}^{36} \left[\frac{H_{\text{obs},i} - H(z_i)}{\sigma_{H,i}}\right]^2,
\end{equation}
where \(H_{\text{obs},i}\) and \(\sigma_{H,i}\) are the observed values and errors, respectively.\\
\textit{\textbf{Combined Constraints}}\\
The total chi-square is given by:
\begin{equation}\label{18}
	\chi^2_{\text{tot}}(\eta,\beta,\gamma,h) = \chi^2_{\text{SN}} + \chi^2_{\text{BAO}}(\eta,\beta,\gamma,h) + \chi^2_{H(z)}(\eta,\beta,\gamma,h).
\end{equation}
Using the combined datasets, the best-fit parameters \(\eta\), \(\beta\), and \(\gamma\) are determined, providing a comprehensive validation of the model.
\section*{ Parameterized Cosmological Modeling and Analysis}
The process of parameterizing kinematical parameters, including Hubble's parameter, and deceleration parameter, is a pivotal step in unraveling the mysteries of the universe's evolution. These parameters serve as the backbone of modern cosmology, providing a quantitative description of the universe's expansion history. Accurate determination of these parameters is crucial for constraining cosmological models, which in turn enables researchers to distinguish between competing theories.\\
Among these parameters, Hubble's parameter ($H(z)$) holds a special significance, as it directly encapsulates the expansion rate of the universe at various redshifts. By parameterizing $H(z)$, researchers can develop a flexible and model-independent framework for describing the expansion history of the universe. This, in turn, allows for the testing of various cosmological models and the constraint of their parameters. Furthermore, the parameterization of $H(z)$ provides a valuable tool for investigating the properties of dark energy, a mysterious component thought to be responsible for the accelerating expansion of the universe. Ultimately, the parameterization of Hubble's parameter emerges as a powerful tool for elucidating the evolution of the universe and constraining cosmological models.
\subsection{Emergent Hubble's parameter}
\begin{figure}[H]
		\centering
		\includegraphics[scale=0.35]{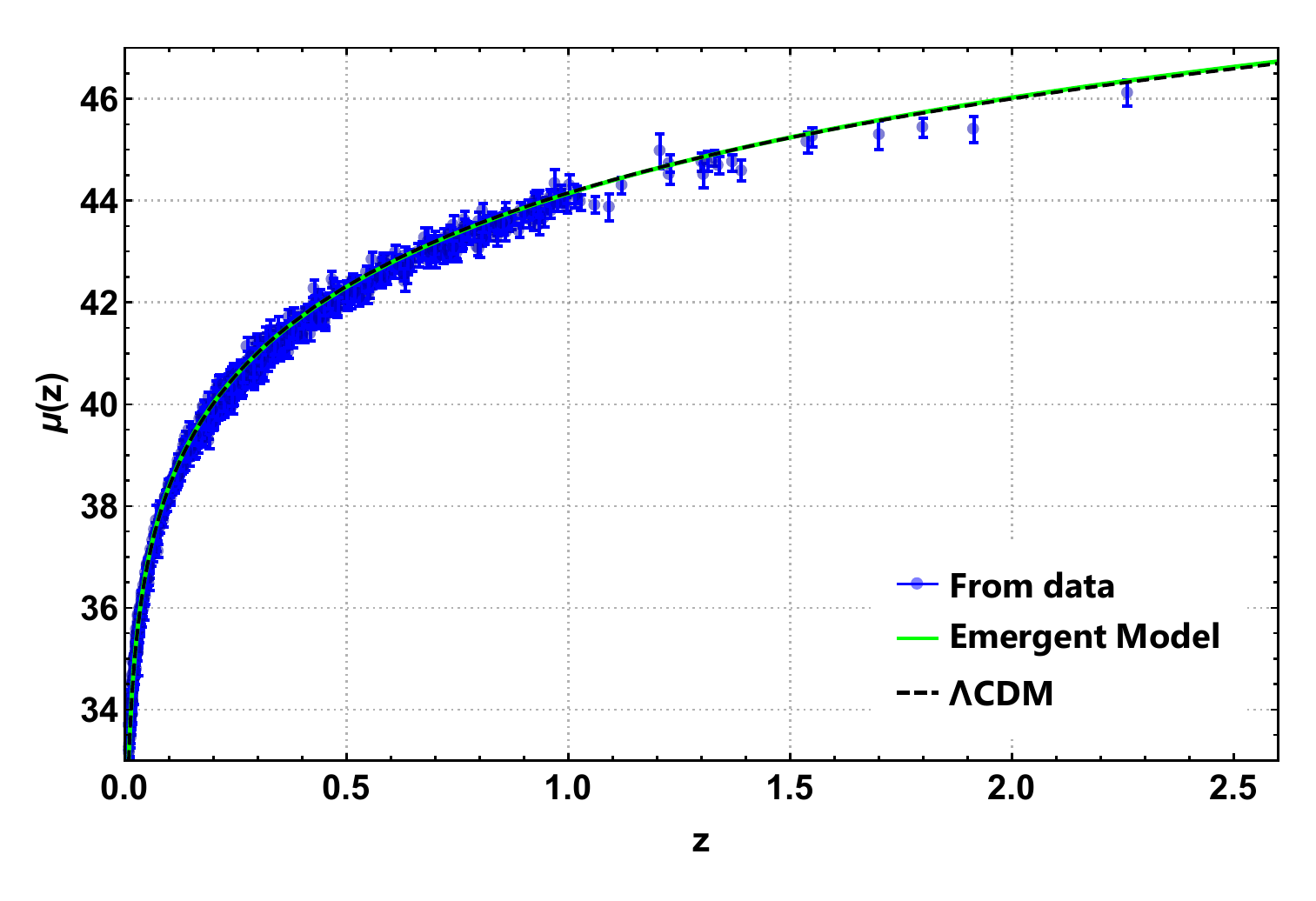}
		\caption{ The behavior of the distance modulus with respect to redshift z is depicted in this graphic. }\label{f1}
\end{figure}

\begin{figure}[H]
		\centering
		\includegraphics[scale=0.35]{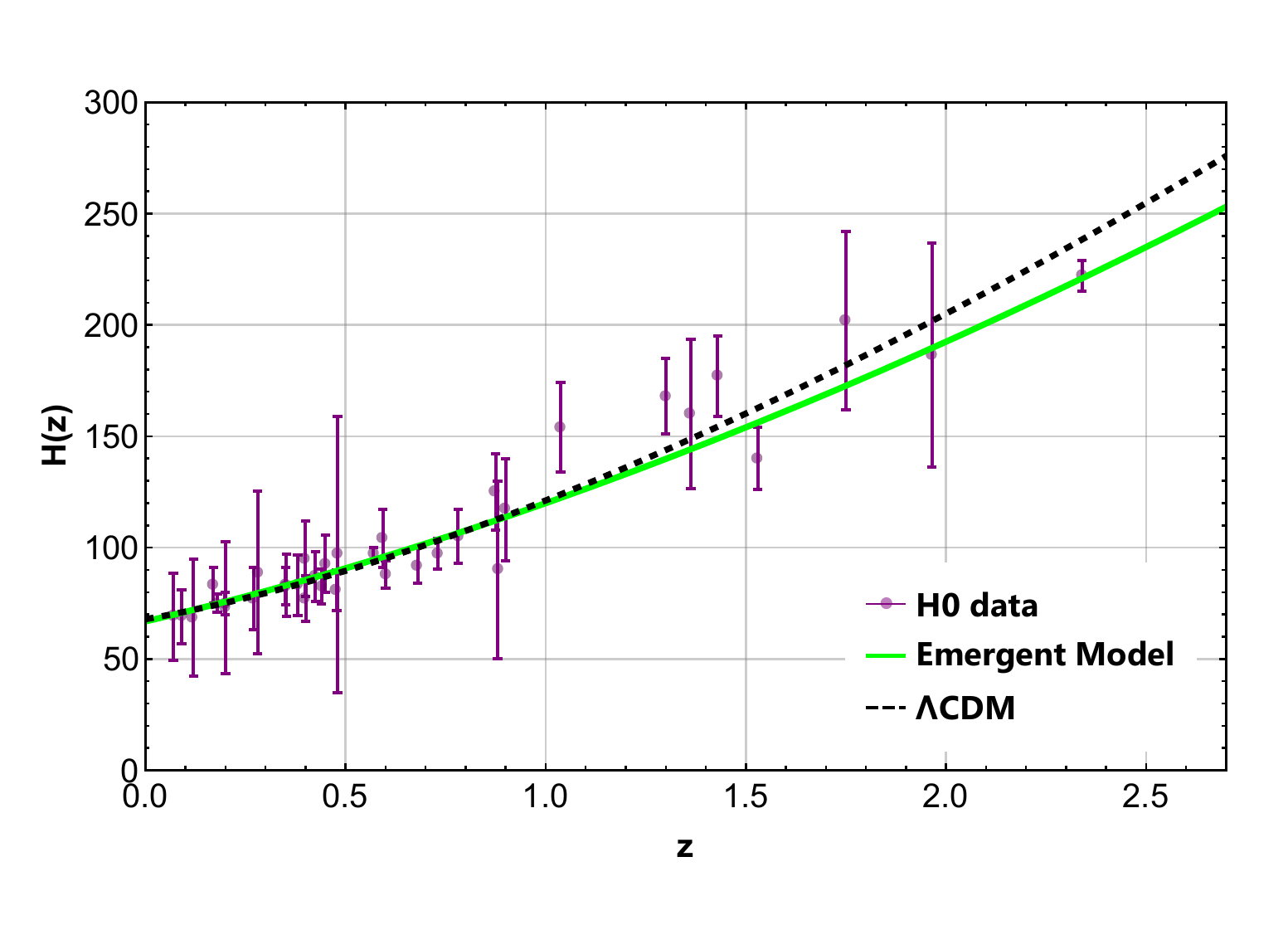}
		\caption{The behavior of the Hubble parameter with respect to redshift z is depicted in this figure. }\label{f2}
\end{figure}

\begin{figure*}
	\centering
	\includegraphics[scale=0.35]{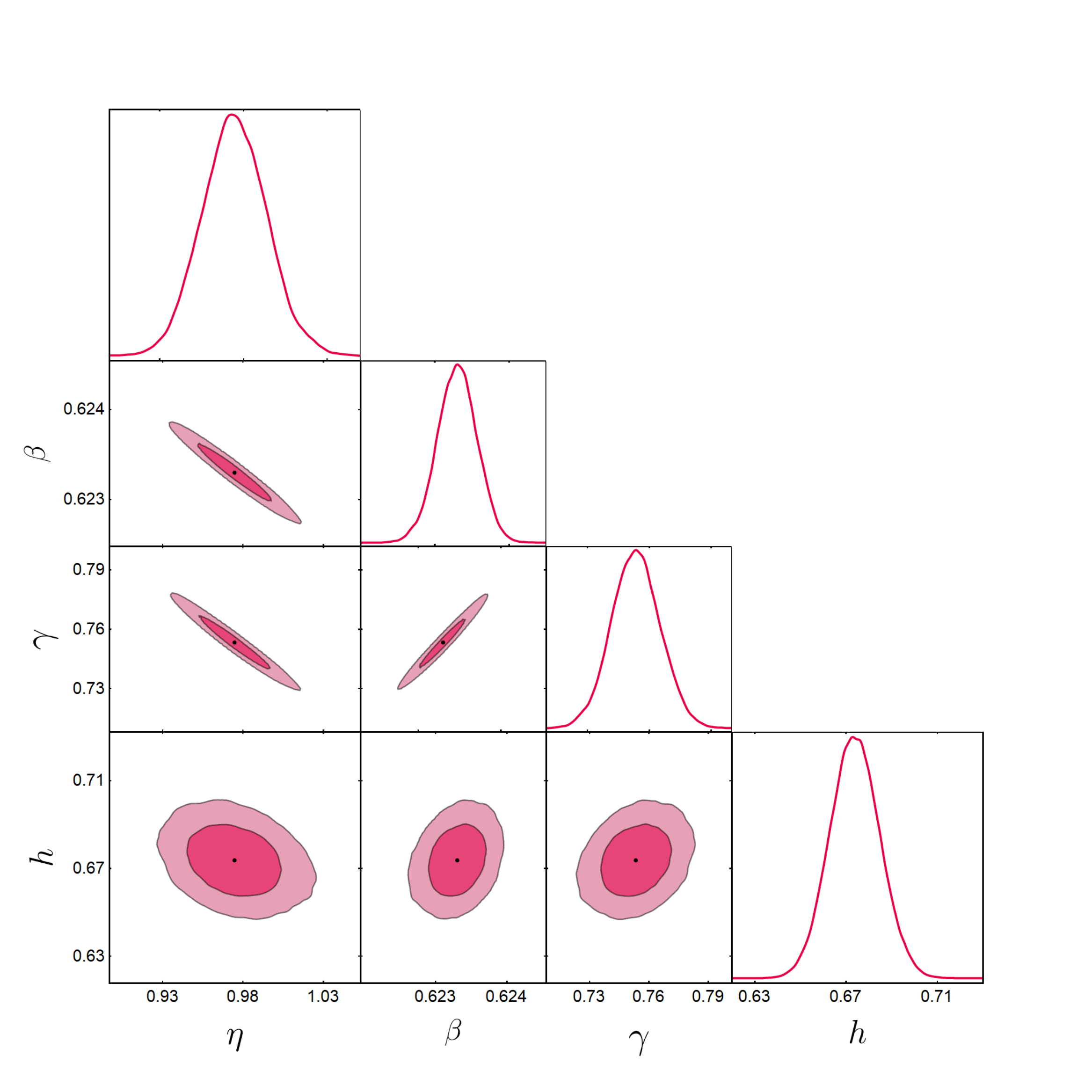}
	\caption{The 1$\sigma$ and 2$\sigma$ confidence contours using the combine datasets for the emergent model of $H(z)$. }\label{contourBR}
\end{figure*}

The emergent universe model presents an intriguing framework with the potential to address some fundamental challenges associated with the standard big bang theory. This model envisions the Einstein static universe as an asymptotic state when traced back in time. From this static phase, the universe transitions into the typical expansion phase, thereby circumventing the big bang singularity. in Ref. \cite{42,43} the authors explored this concept within the modified Starobinsky model, and subsequent research has examined similar scenarios in alternative theoretical frameworks. In its early stages, the model describes a universe large enough to exclude quantum effects. One emergent scale factor, which will be employed in this study, is expressed as follows. We begins with the Hubble parameter of the form redshift as
\begin{equation}\label{19}
	H(z) = H_0 \beta \eta \left[ 1 + \gamma \left( 1 + z \right)^2 \right],
\end{equation}
where \(H_0\) is the Hubble constant, \(\beta\) is a proportionality constant, $\eta$ is a model constant and \(\gamma = -\left( A B^2 \right)^{1/\beta}\). This equation characterizes the variation of the Hubble parameter with redshift \(z\), encapsulating the influence of the emergent scale factor on cosmic expansion. To establish the relationship between cosmic time \(t\) and redshift \(z\), the differential expression $
\frac{dt}{dz} = -\frac{1}{H_0 \beta \left[ 1 + \gamma \left( 1 + z \right)^2 \right] (1+z)}$
is employed. Integration of this expression yields the function
\begin{equation}\label{20}
	t(z) = \frac{1}{H_0 \beta} \log \left[ \frac{1}{\left( A \left( 1 + z \right)^2 \right)^{1/\beta}} \right],
\end{equation}
In this model, \(A\) is a constant associated with the framework. The scale factor is expressed as \(a(t) = A \left[ B + e^{\alpha t} \right]^\beta\), where \(A\), \(B\), and \(\alpha\) are constants chosen to ensure positivity and avoid singularities. For \(\alpha > 0\), the scale factor exhibits exponential growth, reflecting the accelerated expansion of the late universe. The expression in equation (\ref{19}) represents the emergent scale factor of the universe \cite{44}. Originally introduced as a model for a closed system, it describes the transition from an asymptotically static Einsteinian state to one dominated by accelerated expansion. This approach provides a framework for avoiding initial singularities and enables smooth transitions between different cosmic epochs. To constrain the parameters in equation (\ref{19}), we use a combination of observational datasets, including Cosmic Microwave Background (CMB) data from Planck 2018, Supernovae Type Ia (SNIa) data from the Pantheon sample, 36 Hubble parameter measurements from the Hubble dataset, and Baryon Acoustic Oscillation (BAO) data from various surveys 
 The MCMC technique is applied to obtain posterior distributions of the model parameters, with the resulting constraints illustrated through contour plots displaying the 1\(\sigma\) and 2\(\sigma\) confidence intervals (as shown in figures \ref{f1}, \ref{f2} and \ref{contourBR}).

\section{Physical behavior with $H(z)$ parameterization}

Here, we analyze key cosmological parameters and their implications for the evolution of the universe with the considered $H(z)$ Parameterization. This includes investigating the energy density and isotropic pressure, which provide the distribution and dynamics of matter and energy across cosmic scales. The equation of state (EoS) parameter is examined to characterize different cosmic eras, such as matter-dominated, radiation-dominated, and dark energy-dominated phases. Stability of the model is evaluated through the squared sound speed criterion to ensure its physical viability. Additionally, the $\left(\omega-\omega^{\prime}\right)$-plane is explored to study the dynamical behavior of dark energy and its potential transitions between thawing and freezing regions. Finally, we assess the energy conditions—null, dominant, and strong—to ensure the model adheres to fundamental physical principles and to investigate potential deviations from standard General Relativity, further enriching our understanding of cosmic acceleration.
\subsubsection{The energy density}
The expression of energy density for the emergent hubble parameter is obtained as 
\begin{small}
\begin{equation}\label{21}
\rho=\frac{\beta  \eta  H_{0} \left(\beta  \eta  H_{0} (6 \lambda +3) \left(\gamma  (z+1)^2+1\right)^2-4 \gamma  \lambda  (z+1)\right)}{8 \lambda ^2+6 \lambda +1}
\end{equation}
\end{small}
\subsubsection{The isotropic pressure}
The expression of isotropic pressure for the emergent hubble parameter is obtained as 
\begin{small}
\begin{equation}\label{22}
	p=-\frac{\beta  \eta  H_{0} \left(\beta  \eta  H_{0} (6 \lambda +3) \left(\gamma  (z+1)^2+1\right)^2+4 \gamma  (3 \lambda +1) (z+1)\right)}{8 \lambda ^2+6 \lambda +1}
\end{equation}
\end{small}
\begin{figure}[h]
		\centering
		\includegraphics[scale=0.7]{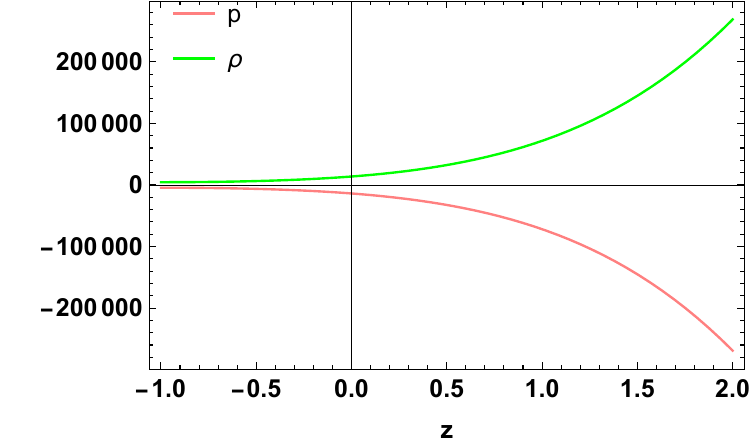}
		\caption{How the variable $\rho$ evolves in comparison to $z$. }\label{pden}
\end{figure}
The density parameter, as shown in Fig. \ref{pden} (\textit{green curve}), remains positive throughout the Universe's evolution and increases with rising redshift \(z\). At high redshifts, it exhibits a significant positive value, gradually approaching zero over time. This behavior is consistent with theoretical expectations, reinforcing the validity of our model. In contrast, Fig. \ref{pden} (\textit{red curve}) reveals an intriguing trend in pressure: it starts from zero at high redshifts and transitions to negative values in the current epoch. This shift to negative pressure reflects the influence of dark energy (DE), which drives the accelerated expansion of the Universe. The observed negative pressure, consistent with the properties of DE, provides empirical evidence for cosmic acceleration and further supports the proposed model. The \(f(R,T) = R + 2\lambda T\) cosmological framework is constrained using recent observational data from a combined dataset, selected for its close alignment with trends observed in other datasets. The joint observational data is employed to examine the evolution of key cosmological parameters, including the density parameter, pressure, deceleration parameter, and effective equation of state (EoS) parameter.
\subsubsection{The equation of state parameter}
The expression of equation of state parameter for the emergent hubble parameter is obtained as 
\begin{small}
\begin{equation}\label{23}
\omega=\frac{4 \gamma  (4 \lambda +1) (z+1)}{4 \gamma  \lambda  (z+1)-3 \beta  \eta  H_{0} (2 \lambda +1) \left(\gamma  (z+1)^2+1\right)^2}
\end{equation}
\end{small}
The evolution of the equation of state (EoS) parameter \(\omega = \frac{p}{\rho}\), as depicted in the figure, plays a crucial role in understanding the properties of the cosmic fluid and its development within the modified \(f(R,T)\) gravity framework with an emergent scale factor. The plot in Fig. \ref{eos} illustrates how \(\omega\) varies with redshift \(z\), covering the universe’s transition from its early stages (\(z > 0\)) to the far future (\(z < 0\)). In the early universe, for \(z > 2\), the EoS parameter remains close to zero, signifying a matter-dominated phase where the pressure \(p\) is negligible compared to the energy density \(\rho\). As the universe evolves towards the present (\(z \approx 0\)), \(\omega\) drops and approaches \(-1\), indicative of a dark energy-like component responsible for the observed accelerated expansion. This shift aligns with current cosmological observations, including Type Ia supernovae and CMB data, suggesting a transition from a decelerating to an accelerating universe. For \(z < 0\), representing the far future, the EoS parameter settles at \(\omega = -1\), corresponding to a de Sitter phase governed by a cosmological constant-like behavior. This implies that the universe will continue to expand at an accelerated pace indefinitely, consistent with predictions of a vacuum energy-dominated scenario. The behavior of \(\omega\) in this framework is consistent with and extends previous studies in modified gravity and dark energy models: 
\begin{figure}[h]
	\centering
	\includegraphics[scale=0.7]{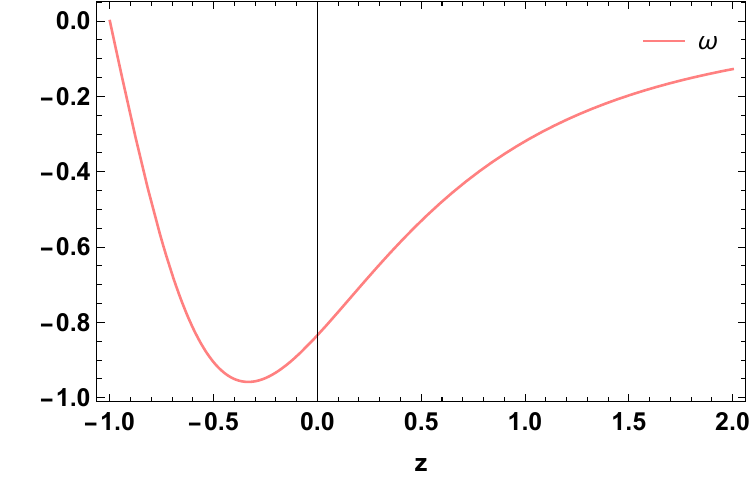}
	\caption{ $\omega$'s behavior in comparison to $z$.}\label{eos}
\end{figure}
\begin{itemize}
	\item In the \(\Lambda\)CDM model, \(\omega = -1\) throughout the dark energy-dominated era, which corresponds to a static cosmological constant. However, the modified \(f(R,T)\) gravity introduces deviations in \(\omega\) during the transition period (\(z > 0\)), reflecting the dynamical nature of dark energy components. 
	\item The EoS parameter's stabilization at \(-1\) in the far future is similar to predictions from quintessence and k-essence models, where \(\omega\) evolves dynamically but eventually mimics a cosmological constant in the late universe.
	\item Comparisons with observational data, such as the Pantheon supernovae dataset and Planck results, show that the current value of \(\omega\) is close to \(-1\), supporting the validity of the modified \(f(R,T)\) gravity model in describing late-time acceleration.
\end{itemize}
The results underscore the effectiveness of the emergent scale factor in modeling the universe's transition from deceleration to acceleration while avoiding singularities. The dynamical nature of \(\omega\) in this model allows for deviations from \(\Lambda\)CDM, offering a richer framework to address unresolved questions in cosmology, including the origin and evolution of dark energy.
\subsubsection{The stability of the model}
\begin{figure}[H]
		\centering
		\includegraphics[scale=0.7]{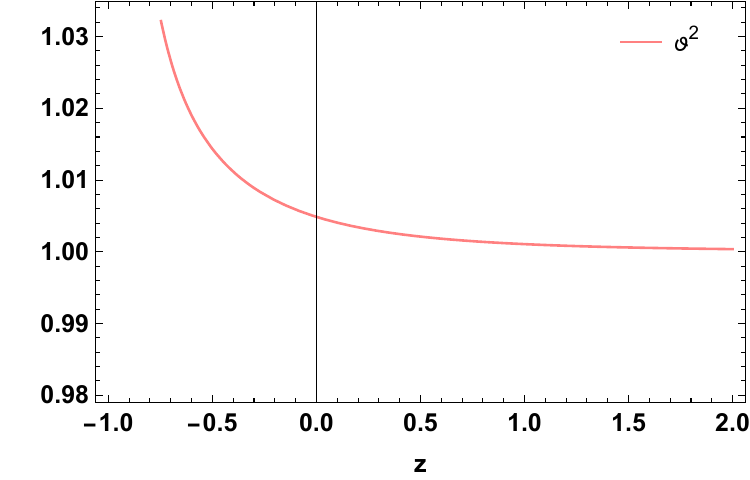}
		\caption{The behavior of $\vartheta^2_s$ versus $z$. }\label{v}
\end{figure}

The expression of stability parameter for the emergent hubble parameter is obtained as 
\begin{equation}\label{24}
	\vartheta^2_s=-\frac{\beta  \eta  H_{0} (6 \lambda +3) (z+1) \left(\gamma  (z+1)^2+1\right)+3 \lambda +1}{\beta  \eta  H_{0} (6 \lambda +3) (z+1) \left(\gamma  (z+1)^2+1\right)-\lambda }
\end{equation}
The squared velocity of sound (\(v_s^2\)) serves as a crucial stability parameter in cosmology, indicating whether perturbations in the cosmic fluid evolve stably or grow uncontrollably over time. The attached Fig. \ref{v} displays the behavior of \(v_s^2\) as a function of redshift (\(z\)) in the context of modified \(f(R,T)\) gravity with an emergent scale factor. The graph reveals that \(v_s^2\) remains positive across all redshift values, signifying the stability of the cosmic fluid throughout the universe's evolution. In the early universe (\(z > 0\)), \(v_s^2\) shows higher positive values, corresponding to a stable matter and radiation-dominated era. This stability ensures that small perturbations in the energy density and pressure dissipate rather than grow, maintaining the universe's homogeneous and isotropic nature during this phase. As the universe transitions toward the present epoch (\(z = 0\)), the values of \(v_s^2\) decrease slightly but remain positive. This behavior reflects the universe's shift from a decelerating expansion dominated by matter to an accelerating expansion dominated by dark energy. The positive \(v_s^2\) in this phase ensures that the perturbations remain under control, aligning with observations of large-scale structure and cosmic microwave background (CMB) anisotropies. In the far future (\(z < 0\)), \(v_s^2\) asymptotically approaches a positive value, indicative of a de Sitter-like phase dominated by a cosmological constant-like dark energy. This stabilization of \(v_s^2\) reflects the universe's transition to a steady-state expansion where perturbations in the cosmic fluid do not lead to instability.

The observed behavior of \(v_s^2\) in this study aligns well with findings in other modified gravity and dark energy models:
\begin{itemize}
	\item In general relativity (GR) and \(\Lambda\)CDM cosmology, \(v_s^2\) is typically positive during dark energy-dominated phases, ensuring the stability of perturbations. The purely positive \(v_s^2\) in this model demonstrates its consistency with the standard framework while introducing modifications specific to \(f(R,T)\) gravity.
	\item Scalar field models such as quintessence often exhibit similar trends in \(v_s^2\), where positivity ensures stability during the transition from matter to dark energy domination. This study's findings further support the robustness of dynamical models in capturing cosmic evolution.
\end{itemize}
The emergent scale factor ensures smooth transitions between cosmic epochs while preserving stability, as evidenced by the positive \(v_s^2\) across all redshifts. This behavior highlights the viability of the modified \(f(R,T)\) gravity framework in describing a stable and consistent evolution of the universe from early times to the far future.
\subsubsection{The $\left(\omega-\omega^{\prime}\right)$ plane}
\begin{figure}[H]
		\centering
		\includegraphics[scale=0.7]{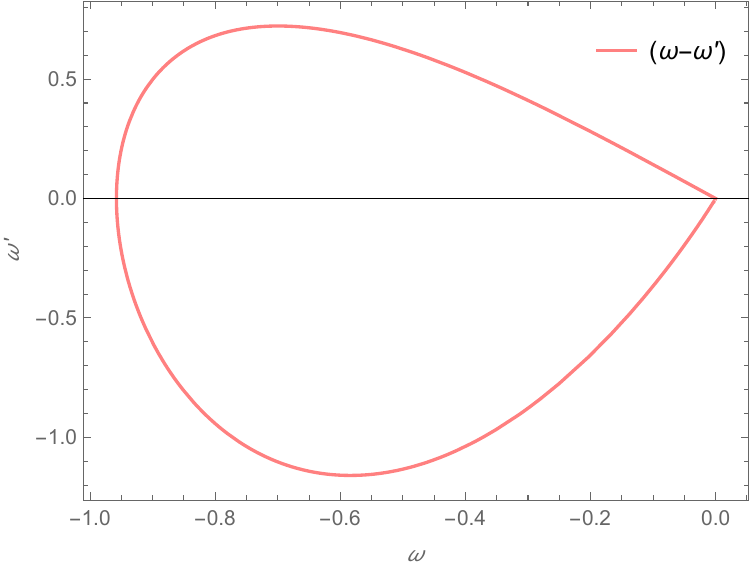}
		\caption{ The behavior of $\omega$ versus $\omega^{\prime}$.}\label{ww}
\end{figure}

The expression of $\omega^{\prime}$ for the emergent hubble parameter is obtained as 
\begin{small}
\begin{equation}\label{25}
	\omega^{\prime}=\frac{12 \beta  \gamma  \eta  H_{0} \left(8 \lambda ^2+6 \lambda +1\right) (-z-1) \left(\gamma  (z+1)^2 \left(3 \gamma  (z+1)^2+2\right)-1\right)}{\left(\beta  \eta  H_{0} (6 \lambda +3) \left(\gamma  (z+1)^2+1\right)^2-4 \gamma  \lambda  (z+1)\right)^2}
\end{equation}
\end{small}
The \((\omega - \omega')\) plane serves as a diagnostic tool to understand the dynamics of dark energy by examining the relationship between the equation of state (EoS) parameter \(\omega\) and its evolution \(\omega' = \frac{d\omega}{d\ln a}\), where \(a\) is the scale factor. The attached Fig. \ref{ww} illustrates the trajectory of the \((\omega - \omega')\) plane in the context of modified \(f(R,T)\) gravity with an emergent scale factor, capturing the dynamical evolution of the universe from early times (\(z > 0\)) to the distant future (\(z < 0\)).

The graph reveals that the trajectory lies predominantly in the region associated with thawing dark energy models. At high redshifts (\(z > 0\)), the EoS parameter \(\omega\) is slightly below zero, indicating a transition from a matter-dominated phase. The corresponding \(\omega'\) values are small but negative, suggesting a gradual evolution of the dark energy component. As the universe evolves toward the present epoch (\(z = 0\)), \(\omega\) approaches \(> 0\), reflecting the dominance of dark energy and its cosmological constant-like behavior. The evolution slows down in this phase, with \(\omega'\) tending toward zero.

In the far future (\(z < 0\)), the trajectory converges toward \((\omega, \omega') = (-1, 0)\), characteristic of a de Sitter phase. This stabilization reflects the asymptotic approach to a vacuum energy-dominated universe, where dark energy mimics a cosmological constant and drives perpetual accelerated expansion.

The observed trajectory in the \((\omega - \omega')\) plane aligns with findings from other studies in modified gravity and dark energy models:
\begin{itemize}
	\item In the \(\Lambda\)CDM model, \(\omega\) remains constant at \(-1\), with \(\omega' = 0\). This study shows that while \(f(R,T)\) gravity mimics this behavior in the distant future, it allows for deviations during the transition phases, reflecting a dynamical dark energy component.
	\item Quintessence models often exhibit thawing behavior, similar to the trajectory observed here, where \(\omega\) starts above \(-1\) and evolves dynamically. 
\end{itemize}
The dynamical evolution observed in the \((\omega - \omega')\) plane underscores the flexibility of the \(f(R,T)\) gravity framework in describing the universe's transition from deceleration to acceleration. The emergent scale factor ensures smooth transitions across epochs while maintaining consistency with observational data. The results emphasize the potential of this model as a viable alternative to \(\Lambda\)CDM.
\subsubsection{The energy conditions}
In cosmology, energy conditions are significant, serving as guidelines to understand the physical behavior of energy and pressure in gravitational theories. The attached Fig. \ref{ec} illustrates the behavior of the null energy condition (NEC), dominant energy condition (DEC), and strong energy condition (SEC) as functions of redshift (\(z\)) in the context of modified \(f(R,T)\) gravity with an emergent scale factor. Below is a detailed physical interpretation of these energy conditions:
\subsection*{1. Null Energy Condition (NEC)}
The NEC is expressed as:
\begin{equation}\label{26}
\rho + p \geq 0,
\end{equation}
where \(\rho\) is the energy density and \(p\) is the pressure. From the graph (\textit{pink curve}), the NEC is satisfied throughout the evolution of the universe, as \(\rho + p\) remains positive across all values of \(z\). At higher redshifts (\(z > 0\)), the NEC shows a significant positive value, reflecting the dominance of matter and radiation in the early universe. As the redshift decreases (\(z \to 0\)), the NEC gradually declines but remains positive, consistent with the transition to a dark energy-dominated phase. In the far future (\(z < 0\)), the NEC asymptotically approaches zero, indicative of a steady-state expansion driven by vacuum energy.
\subsection*{2. Dominant Energy Condition (DEC)}
The DEC is given by:
\begin{equation}\label{27}
\rho \geq |p|.
\end{equation}
This condition ensures that the energy density dominates over the pressure, maintaining the causal structure of spacetime. The graph shows that \(\rho - p\) remains positive for all values of \(z\) (\textit{green curve}), confirming that the DEC is satisfied throughout the universe's evolution. At high redshifts, the positive values of \(\rho - p\) signify the prevalence of matter and radiation, which contribute to the energy density. As the universe transitions to a dark energy-dominated phase (\(z \to 0\)), the DEC remains valid, indicating that the energy density continues to outweigh the effects of negative pressure. In the late universe (\(z < 0\)), the DEC stabilizes, reflecting the dominance of a cosmological constant-like component.
\subsection*{3. Strong Energy Condition (SEC)}
The SEC is expressed as:
\begin{equation}\label{28}
\rho + 3p \geq 0.
\end{equation}
The graph (\textit{red curve}) reveals that the SEC is violated during the late-time accelerated expansion (\(z \leq 0\)), where \(\rho + 3p < 0\). This violation is consistent with the presence of dark energy, which induces repulsive gravitational effects necessary for cosmic acceleration. However, at earlier times (\(z > 0\)), the SEC is satisfied, as the contributions from matter and radiation dominate the pressure term. 

The behavior of the energy conditions in this model aligns with and extends findings in other modified gravity and dark energy studies:
\begin{itemize}
	\item In standard general relativity (GR) and \(\Lambda\)CDM models, the NEC and DEC are typically satisfied, while the SEC is violated during late-time acceleration. The results here are consistent with these trends while introducing additional features specific to \(f(R,T)\) gravity.
	\item Scalar field models, such as quintessence and k-essence, exhibit similar behavior, where the SEC is violated during dark energy domination, reflecting the universal nature of this condition in explaining late-time acceleration.
\end{itemize}
The analysis of energy conditions highlights the consistency of \(f(R,T)\) gravity with observational constraints and its ability to describe the universe's transition from deceleration to acceleration. The satisfaction of NEC and DEC ensures the physical viability of the model, while the violation of SEC underscores the necessity of modifications to GR to account for late-time cosmic acceleration. The emergent scale factor plays a key role in shaping these behaviors, providing smooth transitions between different epochs.
\begin{figure}[h]
	\centering
	\includegraphics[scale=0.7]{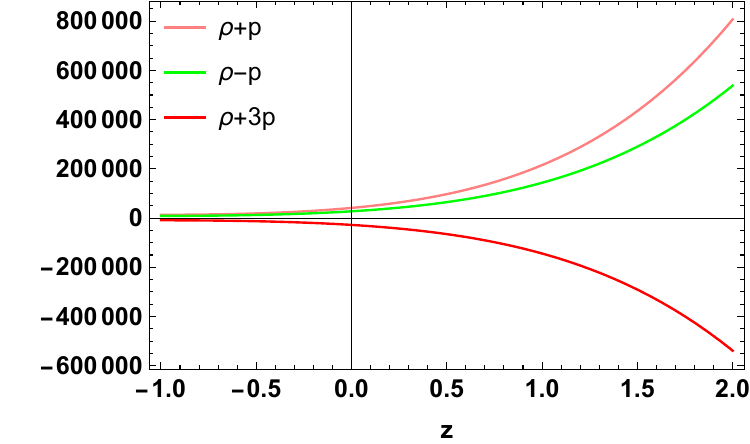}
	\caption{The behavior of $\rho+p$, $\rho-p$ and $\rho+3p$ versus $z$. }\label{ec}
\end{figure}
\section{Physical Interpretation of the Parameters in $f(R,T)$ Gravity}
In this study, we investigate the modified $f(R,T)$ gravity framework in the context of FRW spacetime using an emergent scale factor. The analysis focuses on key cosmological parameters such as ED, isotropic pressure, EoS parameter, stability of the model, $(\omega - \omega')$ plane behavior, and energy conditions. Figs. \ref{pden} to \ref{ec} present the graphical behavior of these parameters, which are explained in detail below.
	
	The {\bf energy density} decreases monotonically from the early universe ($z > 0$) to the late universe ($z < 0$). At high redshifts, the energy density dominates, corresponding to the radiation and matter-dominated eras. This aligns with the standard cosmological models, where the energy density decreases due to the expansion of the universe. The emergent scale factor, however, introduces deviations, particularly in the transition from deceleration to acceleration. These deviations highlight the role of $f(R,T)$ modifications, where the coupling between the Ricci scalar $R$ and the trace of the energy-momentum tensor $T$ influences the cosmic dynamics.
	
	The \textbf{isotropic pressure} remains negative throughout cosmic evolution, becoming less negative as the universe transitions from the early to the late phases. Negative pressure is characteristic of dark energy, and its behavior here suggests that the $f(R,T)$ gravity model effectively captures late-time acceleration. The increase in $p$ (less negative values) as $z \to -1$ indicates the universe approaching a de Sitter-like phase, driven by dark energy dominance.
	
	The \textbf{equation of state parameter} Initially, $\omega$ is near zero, corresponding to a matter-dominated phase. It then decreases through $-1$, representing a cosmological constant-like dark energy, and asymptotically stabilizes near $-1$ in the far future. This trajectory reflects the transition from deceleration to acceleration, a hallmark of late-time cosmic expansion. The small deviations of $\omega$ from $-1$ highlight the dynamical nature of dark energy in $f(R,T)$ gravity, distinguishing it from the cosmological constant in $\Lambda$CDM.
	
	The \textbf{stability parameter} The stability is ensured when $v_s^2 > 0$. The graph shows that the model is stable across cosmic evolution, with minor regions of instability near transition phases. This result confirms that the modifications in $f(R,T)$ gravity allow for a stable late-time accelerated expansion.
	
	The \textbf{$(\omega - \omega')$ plane}, shows a thawing behavior, where $\omega$ evolves from values near $-1$ with a small derivative $\omega'$. This trajectory reflects the dynamical nature of dark energy in $f(R,T)$ gravity, contrasting with the constant $\omega = -1$ of $\Lambda$CDM. The slope and curvature of the trajectory in the $(\omega - \omega')$ plane further emphasize the model's flexibility in describing the universe's acceleration.
	
	Lastly, the \textbf{energy conditions} (NEC, SEC, DEC, and WEC) are critical for assessing the model's physical viability. The graphical analysis reveals that some conditions, such as SEC, are violated, while others, such as NEC and WEC, are satisfied. The violation of SEC supports the accelerated expansion, while the satisfaction of NEC ensures causal consistency. These results validate the $f(R,T)$ model's ability to describe a universe transitioning from deceleration to acceleration, capturing both early and late-time behaviors.

	\section{Conclusion}
	In conclusion, this study explores the modified $f(R,T)$ gravity framework in the context of FRW spacetime with an emergent scale factor, providing a detailed analysis of key cosmological parameters. The energy density, isotropic pressure, equation of state parameter, stability, and energy conditions all reflect the model's capability to describe the universe's evolution, particularly its transition from deceleration to acceleration. The emergent scale factor introduces notable deviations in the cosmic dynamics, particularly during the shift to accelerated expansion, highlighting the influence of $f(R,T)$ modifications. The model captures late-time acceleration effectively, with the isotropic pressure and EoS parameter indicating dark energy dominance, and the stability analysis ensuring the model's physical viability. Additionally, the behavior in the $(\omega - \omega')$ plane and the energy conditions support the model's ability to describe a universe transitioning from deceleration to acceleration, making it a promising framework for understanding cosmic evolution in the thawing region.

\section*{Declaration of competing interest}
The authors declare that they have no known competing interests.

\section*{Data availability}
The research presented in the paper did not use any data.

\section*{Acknowledgments}
The authors, S. H. Shekh, A. Pradhan, and A. Dixit, thank the IUCAA, Pune, India, for providing the facility through the Visiting Associateship programs. Additionally, the Science Committee of the Republic of Kazakhstan's Ministry of Science and Higher Education provided funding for this study (Grant No. AP23483654).

\end{document}